\documentclass[apjl]{emulateapj}

\usepackage{amsmath}
\usepackage{graphicx}
\usepackage{epsfig,psfrag,epic,eepic}
\usepackage{textcomp}
\usepackage{times}

\slugcomment{Submitted to ApJ Letters}

\shorttitle{The velocity function of halos at 20 kpc}
\shortauthors{Weinmann et al.}

\begin{document}

\title{The velocity function of dark matter halos at $r=20$
  kpc: Remarkably little evolution since $\lowercase{z} \approx$ 4}

\author{Simone M. Weinmann\altaffilmark{1}, Marijn Franx\altaffilmark{1}, Pieter van Dokkum\altaffilmark{2}, Rachel
  Bezanson\altaffilmark{2}}
\email{weinmann@strw.leidenuniv.nl}
\altaffiltext{1}{Leiden Observatory, Leiden University, P.O. Box 9513, 2300 RA
Leiden, The Netherlands}
\altaffiltext{2}{Department of Astronomy, Yale University, P.O. Box 208101, New Haven, CT 06520-8101, USA}

\begin{abstract}\noindent
We investigate the evolution in the dark matter halo circular velocity
function,  measured at a  fixed physical  radius of  20 kpc ($v_{20}$), which is
likely  to  be a  good  proxy for  galaxy  circular  velocity, in  the
Millennium-II simulation.  We find that the $v_{20}$ function
evolves remarkably little since z $\approx$ 4. We analyze the histories of the main progenitors of halos, and
 we find that the dark matter distribution 
within the central 20 kpc of massive halos has been in place since early times. This provides evidence for the inside-out growth of haloes. 
The  constancy of
the central circular velocity of halos may offer a natural explanation
for the observational finding that  the galaxy circular velocity is an
excellent predictor  of various  galaxy properties.  Our  results also
indicate that we can expect a significant number of galaxies with high
circular  velocities  already at  $z=4$  (more  than  one per  $10^{6}
h^{-3}{\rm  Mpc}^{3}$ with circular  velocities in  excess of  450 km
${\rm s}^{-1}$,  and more  than one per  $10^{4.5}h^{-3}{\rm Mpc}^{3}$
with circular velocities in excess of 350 km ${\rm s}^{-1}$).
Finally, adding baryonic mass and  using a simple model
for halo adiabatic contraction,  we find remarkable agreement with the
velocity dispersion functions  inferred observationally by Bezanson et
al. (2011) up to $z \approx 1$ and down to about 220 km ${\rm s}^{-1}$.
\end{abstract}





\section{Introduction}
\label{intro}
Simulations of individual dark matter halos have provided evidence for
the inside-out growth of halos in a $\Lambda$CDM cosmological
model. These studies have shown that the inner 
few tens of kpc remain almost unchanged since high redshift
both in massive halos (e.g. Loeb \& Peebles 2003; 
Gao et al. 2004) and in halos expected to host Milky-Way sized galaxies (Wechsler et al. 2002; Diemand et al. 2007; Wang et al. 2011). From studying the mass-concentration relation of halos in cosmological simulations, it has become clear that the central density of haloes is correlated with their epoch of formation, which provides additional evidence for inside-out growth (Bullock et al. 2001a; Wechsler et al. 2002; Zhao et al. 2003).

The almost constant central densities of halos may have direct consequences for galaxy formation: At $z=0$, the circular velocity 
at 10 kpc for dark matter halos has been found to be a good proxy for
galaxy circular velocity and galaxy velocity
dispersion (Loeb \& Peebles 2003; Trujillo-Gomez et al. 2011). It is
likely that this is also the case at higher redshift, and that central densities of halos can be used to predict galaxy circular velocities.

Motivated by these findings, we study 
the evolution of the circular velocity of dark matter halos at 20 kpc 
(hereafter $v_{20}$) in this work. We choose a radius of 20 kpc because galaxies in the nearby Universe are thought
to have nearly isothermal mass profiles out to this scale. The scale is also small enough
that we can expect structure to have formed very early (e.g. Diemand et al. 2007), and at the same
time large enough that it is not completely dominated by baryonic
processes. 

We extend previous studies, which have focussed on individual halos,
by studying the same quantity
for a statistical sample of dark matter halos in the
Millennium-II high-resolution cosmological simulation. Thus, we can
for the first time explore the evolution of the small-scale circular
velocity function of halos
as a function of redshift. 

\begin{figure*}[t]
\centering
\includegraphics[width=400pt]{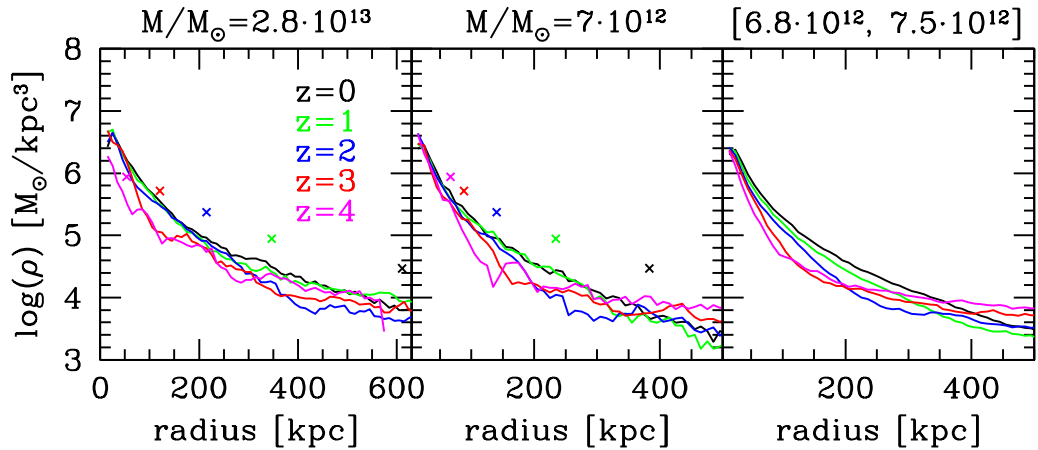}
\caption{Evolution of the density profile of a randomly selected 
halo with a ($z=0$) mass of $7 \cdot
10^{12} M_{\odot}$, and with a virial radius of 370 kpc (left) and 
with a mass of $2.8 \cdot
10^{13} M_{\odot}$ and 610 kpc (middle panel). Crosses show the location of the virial radius at each redshift. In the right panel, we show the 
average profile of 22 halos in a narrow bin in halo mass from  $6.8 \cdot
10^{12} M_{\odot}$ to $7.5 \cdot
10^{12} M_{\odot}$, following the main progenitor history of halos. This figure shows that the density profile
within $\approx$ 20--50 kpc undergoes only little evolution since $z \approx 4$.
}
\label{fig:example}
\end{figure*}

\section{Data}
\label{sec:method}
We use the publicly available dark matter 
data from the Millennium-II simulation (hereafter MS-II, Boylan-Kolchin et al. 2009),
which is contained in a box of side 137 Mpc and has a particle mass of $9.45 \cdot 10^{6} {\rm M_{\odot}}$.
According to eq. 20 in Power et al. (2003) the MS-II mass resolution is sufficient to resolve 
the circular velocity profile of halos down to 20 kpc and $R_{\rm max}$ for 
halos with masses above $10^{10} {\rm M_{\odot}}$. 
Particle data for this simulation is not available. Thus, we calculate $v_{20}$ from $R_{\rm max}$, $R_{200}$ and $v_{200}$ analytically, assuming a Navarro, Frenk \& White (1997, NFW) profile:
\begin{equation}
\rho(R)=\frac{\rho_0}{(R/R_{\rm s})(1+R/R_{\rm s})^2}
\end{equation}
with the scale radius $R_{\rm s}$ and $\rho_0$ as the free parameters of the profile.
The concentration of a NFW halo is defined as the ratio between
  scale radius and virial radius, $c = R_{\rm s}/R_{\rm 200}$. Given
  that the maximum circular velocity is reached at a radius $R_{\rm
    max}=2.163 \cdot R_{\rm s}$ for a NFW halo, it follows that
\begin{equation}
c=2.163 \cdot R_{\rm 200}/R_{\rm max}.
\end{equation}
Integrating the NFW profile, the enclosed mass at radius $R$ is given by
\begin{equation}
M(R)=4\pi\rho_0R_{\rm s}^3\left[\rm ln(1+{{\it cx})-\frac{{\it cx}}{1+{\it cx}}}\right]
\end{equation}
with $x=R/R_{\rm 200}$. Given that $v_{\rm circ}(R) = \sqrt{\frac{GM(<R)}{R}}$ we can then calculate the circular velocity at 20 kpc using
\begin{equation}
v_{\rm 20}=v_{\rm 200} \cdot \left[ \frac{1}{x} \cdot \frac{{\rm ln}(1+cx)-\frac{cx}{1+cx}}{{\rm ln}(1+c)-\frac{c}{1+c}} \right] ^{0.5}
\end{equation}
with $x$=20 kpc/$R_{\rm vir}$.
Klypin et al. (2001) show that NFW profiles fit numerically simulated profiles with a relative error of below about 10 \% for radii larger than 1 \% of the virial radius. The density at 20 kpc should thus be well described by a NFW profile.

For most of our analysis (with the exception of Fig. \ref{fig:example} and \ref{fig:history} where we only consider centrals), we use both central and type 1 satellite galaxies, only
leaving out the 'orphan' galaxies without associated subhalos. For the satellite galaxies, we use $R_{\rm max}$ and  $v_{\rm 200}$ at the time of infall to 
calculate  $v_{20}$, i.e. we assume that $v_{20}$ stays unchanged
after infall. This assumption seems reasonable, see Diemand et al. (2007). 

We also make use of the lower-resolution Milli-Millennium simulation (hereafter Milli-MSI, Springel et al. 2005), 
which has a particle mass of $1.18 \cdot 10^{9} {\rm M_{\odot}}$. For this simulation, individual 
particle data are available, which we use to calculate individual halo profiles, and to
compare with the results from the MS-II simulation. 
Both the MS and MS-II simulations are based on a WMAP1 cosmology (Spergel et al. 2003).

\section{Results}
\subsection{Particle data}
As a first step, we consider the dark matter profiles of central halos in the Milli-MSI 
simulation, calculated from particle data. In Fig. \ref{fig:example},
left and middle panel, we
show the evolution of the density profile up 
to $z=4$ for two 
randomly selected halos. The halo shown in the left hand panel  reaches a virial mass of $2.8 \cdot
10^{13} M_{\odot}$ and a virial radius of 610 kpc, the one in
the middle panel a mass of $7 \cdot
10^{12} M_{\odot}$ and a virial radius of 370 kpc by $z=0$. In the right panel of the same figure, we show the 
average profile of 22 halos in a narrow bin in stellar mass from  $6.8 \cdot
10^{12} M_{\odot}$ to $7.5 \cdot
10^{12} M_{\odot}$. The central parts of haloes (inside 20--50 kpc) show only relatively little change, while there is significant growth on larger scales. We also show the location of the virial radius and
virial density at different redshifts as crosses to indicate the significant evolution of those quantities. As a side note,
absence of change in the inner part of the profile does not
mean that the same dark matter particles stay in the inner region of
the halo since $z \approx 4$ (e.g. Gao et al. 2004).

\begin{figure*}[t]
\centering
\includegraphics[width=350pt]{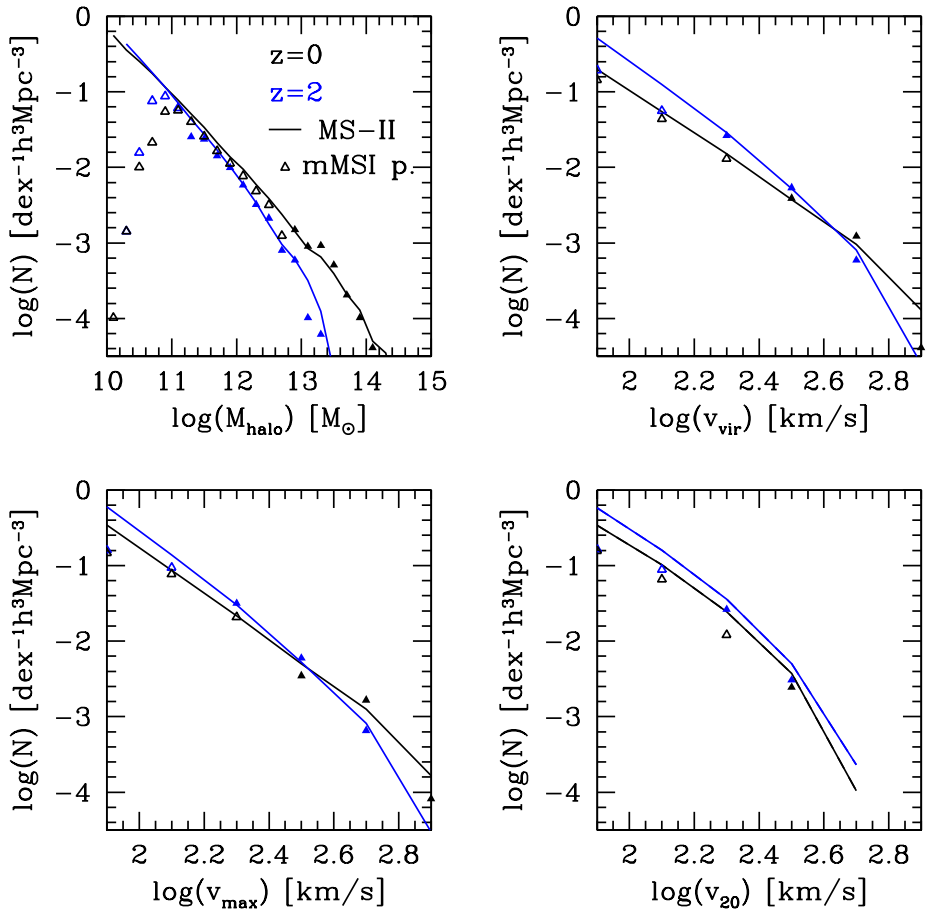}
\caption{We compare results from MS-II (lines) to results
from the Milli-MSI simulation particle data (symbols) down to the 
resolution limit. Filled symbols are results above the resolution limit
according to Power et al. (2003), 
empty symbols below. The similarity of the results shows that velocity profiles
of halos are reasonably well converged even below the
resolution limit by Power et al. (2003).
}
\label{fig:images}
\end{figure*}

As a next step, we check the agreement between the statistical distribution of 
halo properties in the MS-II simulation with results from the Milli-MSI simulation. 
In Fig. \ref{fig:images}, we show the halo mass, virial velocity, maximum circular velocity ($v_{\rm max}$) and $v_{20}$ distribution for both simulations, with  $v_{20}$ calculated directly from the particle data for
the Milli-MSI, and indirectly for the MS-II, at $z=0$ and
$z=2$.
In principle, the inner profile of halos can only be expected to be well resolved 
down to the limit given by Power et 
al. (2003). If we demand that less than 10 \% of Milli-MSI halos fall below this limit,
 the resolution limit is about  $v_{\rm vir}$=250 km ${\rm s}^{-1}$ at $z=0$, and
 $v_{\rm vir}$=117 km ${\rm s}^{-1}$ at $z=2$. 
We find that $v_{\rm 20}$ for the two simulations is in reasonable agreement, although it appears to be 
slightly lower in the Milli-MSI simulation.
\subsection{Evolution in $v_{20}$}
\begin{figure*}[t]
\centering
\includegraphics[width=350pt]{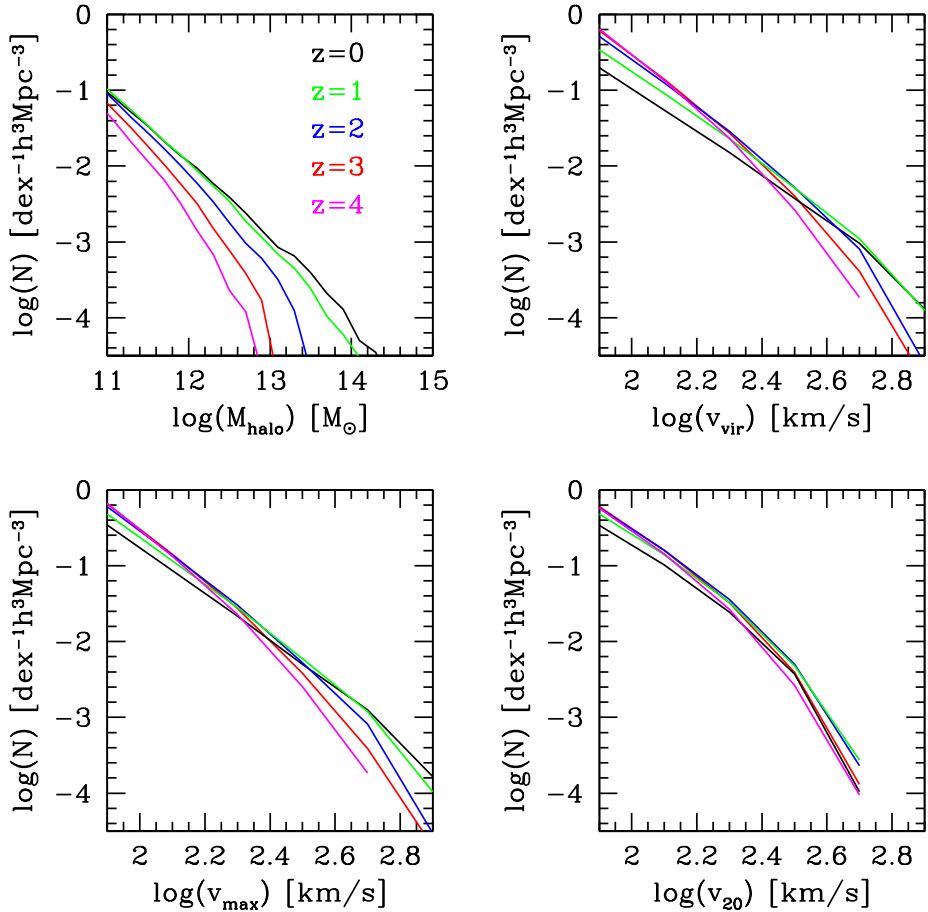}
\caption{Evolution of the halo mass, virial velocity, $v_{\rm max}$ and
$v_{20}$ functions at different redshifts. While the halo mass function
undergoes substantial growth at all masses, the $v_{\rm max}$ and
virial velocity function only grow at the high velocity end, and
actually decrease at the low velocity end. The $v_{20}$ function undergoes remarkably little change since $z=4$.
}
\label{fig:ev_func}
\end{figure*}

\begin{figure*}[t]
\centering
\includegraphics[width=380pt]{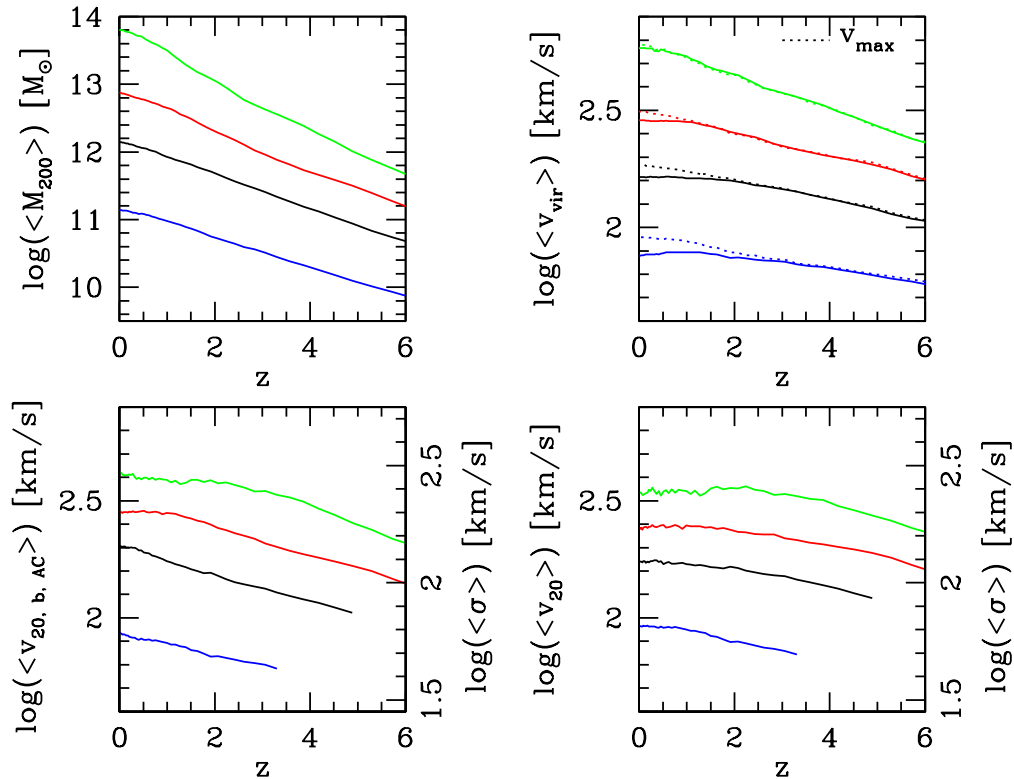}
\caption{Evolution of the average halo mass, virial velocity and $v_{\rm max}$, and 
$v_{20}$, for the main progenitors of halos in four bins of $M_{\rm
  200}$ at $z=0$. The top two panels and the bottom right are based on dark matter results only, the bottom left panel shows $v_{20}$ after correction for baryons and AC. Results for $v_{20}$ are only shown as long as $v_{20}
< v_{\rm max}$. The right axis of the bottom two panels show rescaled results for $\sigma=v_{20}/\sqrt 2$. 
Confirming results from Fig. \ref{fig:ev_func},
$v_{20}$ undergoes only little change since  $z \approx 2-4$, depending on the
halo mass bin.
}
\label{fig:history}
\vspace*{2mm}
\end{figure*}

In Fig. \ref{fig:ev_func} we show the evolution of the halo mass function, 
virial velocity function, $v_{\rm max}$ function and 
$v_{20}$ function. The halo mass function evolves dramatically over this time period. 
The virial velocity function and the $v_{\rm max}$ function also show significant evolution at the high
mass end, indicating the formation of groups and clusters of galaxies. At intermediate values 
of $v_{\rm max} \approx$ 200 km/s, little evolution is seen, in good agreement with results by Bullock et al. (2001b). Finally, very little evolution in the entire range of velocities
is seen for the $v_{20}$ function, which is well in place since
$z=4$. As we show below, 
this is likely connected to the inside-out growth of halos.

We then consider the main progenitor histories of halos in narrow halo mass
bins at $z=0$ in Fig. \ref{fig:history}, averaging over 100--400 central
halos in the three lower bins, and over 43 halos in the highest mass bin. In agreement with the previous figure, we can see
 circular velocities growing more slowly than halo
mass, particularly at late times. Since $z=1$, neither the average 
$v_{20}$ nor the average $v_{\rm vir}$ has changed significantly in all 
mass bins. For the three higher bins, $v_{20}$ has even stayed practically constant since $z=2$. 
This figure is interesting since it allows us to directly read off the expected value
of $v_{20}$ for the main progenitor of a halo with a given final mass at $z=0$. A halo 
with a mass of $ 
10^{13} M_{\odot}$, which is expected to host a galaxy with a stellar mass of about
$ 10^{11} M_{\odot}$ today (e.g. Moster et al. 2012) will have had $v_{20} \approx 250$ km ${\rm s}^{-1}$ since $z=2$, and had reached
 $v_{20} \approx 160$ km ${\rm s}^{-1}$ already at $z=6$. A halo 
with a mass of $ 
10^{12} M_{\odot}$, which is expected to host a galaxy with a stellar mass of about
$2.5 \cdot 10^{10} M_{\odot}$ at $z=0$ will have had $v_{20} \approx 180$ km ${\rm s}^{-1}$ since $z=2$, and $v_{20} \approx 100$ km ${\rm s}^{-1}$ already at $z=6$. 
From this we conclude that the main progenitors of today's massive
galaxies will likely exhibit high circular velocity already at high 
redshifts. 

\subsection{Comparison to observational results}

As a next step, we compare our findings to observational results of the galaxy velocity dispersion $\sigma$ by Bezanson et al. (2011)
and Bernardi et al. (2010), assuming  $\sigma$= $v_{20}/\sqrt2$. Bernardi et al. (2010) calculate velocity dispersion functions
from the SDSS using the spectroscopically measured quantity within the SDSS fiber. Bezanson et al. (2011) estimate the velocity dispersion of higher redshift galaxies using the stellar masses
and radii, using a method calibrated on $z=0$ data.

\begin{figure*}[t]
\centering
\includegraphics[width=500pt]{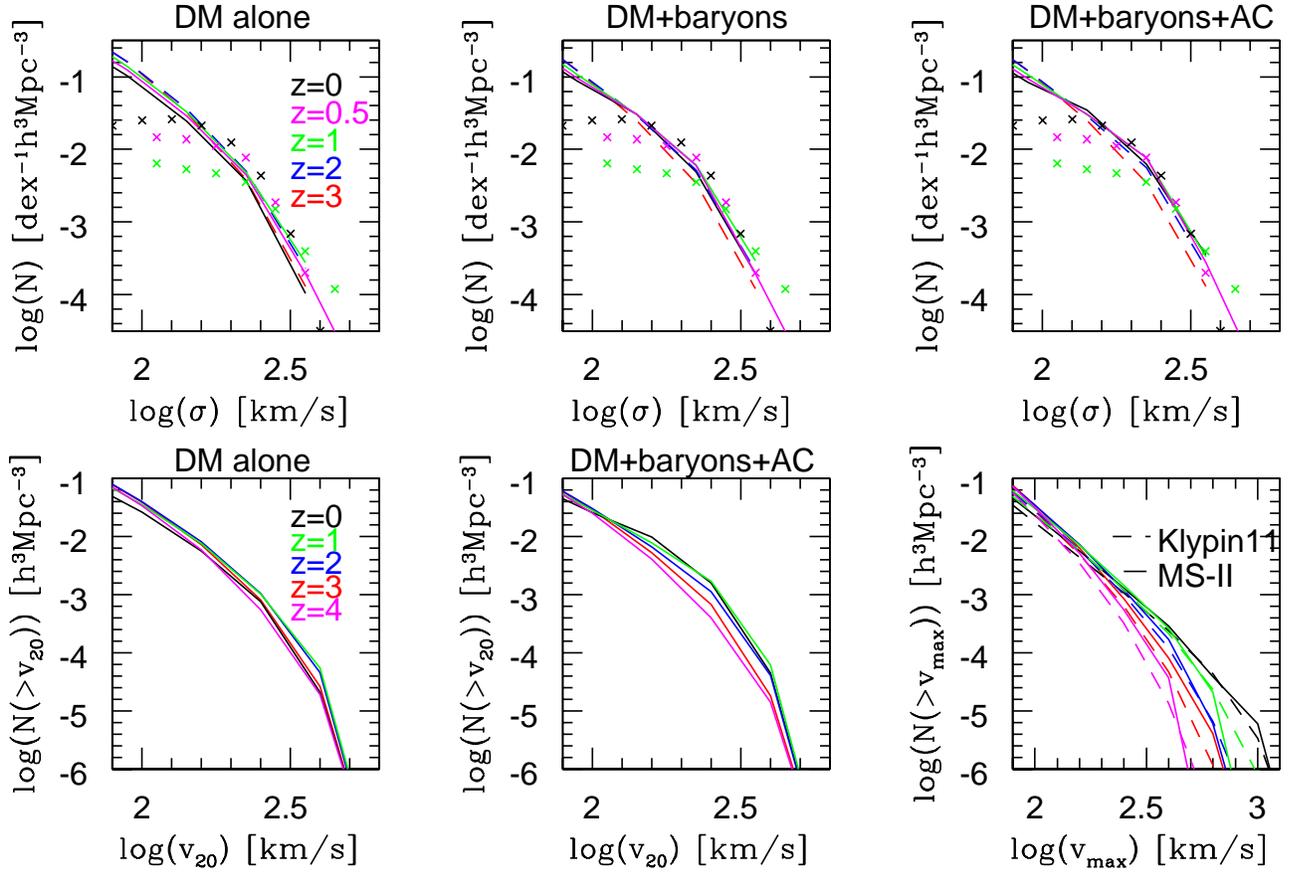}
\caption{Top panels: Comparison of $\sigma$ at 20 kpc (i.e. $v_{20}/\sqrt2$) to results by 
Bezanson et al. at $z=0.45$ and $z=1.05$, and Bernardi et al. (2010) at $z=0$ (crosses). 
We also show predictions for $z=2$ and $z=3$ as dashed lines. Results are shown for dark matter alone (top left panel), dark matter and baryons (top middle panel), and dark matter, baryons and AC (top right panel). Agreement
between theoretical predictions and the Bezanson et al. (2011) results are
excellent for the high mass end of the velocity dispersion function.
Bottom panels: Cumulative functions. Bottom left panel: $v_{\hspace{0mm}20}$-function, only taking into account dark matter. Middle panel: Same like left hand panel, but with the contribution of baryons and AC added.
 Right hand panel: Cumulative $v_{\rm max}$-function, compared to the analytical fits of Klypin et al. (2011) based on the Bolshoi simulation.}
\label{fig:rachel}
\end{figure*}

Assuming no halo contraction and no baryonic content, the agreement between the MS-II results
and the observations is already 
encouraging (Fig. \ref{fig:rachel}, top left panel). But of course, the presence of 
baryonic matter in the central regions of the halo, and 
the potential effect of halo contraction (Blumenthal et al. 1986; Trujillo-Gomez et al. 2011)
should both serve to increase the central circular velocity dispersion.  

We add baryonic mass according to
\begin{equation}
M_{\rm 20 kpc} = M_{\rm 20 kpc, DM}\cdot(1-f_{\rm b}) + M_{*}\cdot(1+f_{\rm gas})
\end{equation}
with $f_{\rm b}=0.17$ the cosmic baryon fraction.
We estimate $M_{*}$ from the redshift-dependent relation between halo mass and stellar mass as given 
in Moster et al. (2012). For the cold gas fraction, $f_{\rm gas} = M_{\rm gas}/M_{*}$, we use the average
$f_{\rm gas}(z, M_{*})$ from the Guo et al. (2011) semi-analytical model. The result is shown in the top middle panel of  Fig. \ref{fig:rachel}. The difference to the dark matter-only  model is small.

As a next step, we take into account the effect of adiabatic contraction (AC) using
the Blumenthal et al. (1986) formula, iteratively solving the equation
\begin{equation}
r_{\rm i}\cdot M_{\rm tot}(r_{\rm i})= r_{\rm f}\cdot [M_{\rm tot}(r_{\rm i})\cdot(1-f_b) + M_{*}\cdot(1+f_{\rm gas})]
\end{equation}
for $r_{\rm i}$, with $r_{\rm f}=$20 kpc.
In case this formula results in halo expansion instead of contraction, we
assume that the halo profile simply stays fixed. While the magnitude of the effect
of adiabatic contraction, and even its existence, is still unclear and might differ for different galaxy types (e.g. Gnedin et al. 2004; 
Dutton et al. 2011), taking into account both the contribution of baryonic material and AC results
 in near perfect agreement with the Bezanson result (top right panel 
of Fig. \ref{fig:rachel}) at the high velocity dispersion end. At lower velocity dispersions, the agreement breaks down. This is similar to the results obtained by Trujillo-Gomez et al. (2011) at $z=0$.
 
The disagreement between our predictions and the results of Bezanson et al. (2011) at lower velocity dispersions may indicate that the simple relation between velocity dispersion and circular velocity breaks down for disky galaxies with 
a small bulge component. In addition, the estimates for velocity dispersions used by Bezanson et al. (2011)
might become progressively uncertain for lower-mass galaxies, due to the increasing
contribution of face-on disks and the higher gas and dark matter content of low-mass galaxies (but see Taylor et al. 2010 who find that the method works well down to 100 km ${\rm s}^{-1}$ at $z=0$). The fact that the agreement improves  towards lower redshifts might indicate that velocity dispersions are particularly difficult to estimate for highly star-forming, gas-rich galaxies. Bezanson et al. (2012) show that the $\sigma$ below which more than 50 \% of all galaxies are star-forming decreases towards lower redshift, a signature of downsizing. In all, the cause of the disagreement at low velocity dispersions is not entirely clear, but the agreement is excellent in the high-mass regime where elliptical, passive and probably gas-free galaxies dominate.
We note that previous studies have found
strong and puzzling 
disagreement between observations and model predictions at even lower velocity dispersions, below 100 km ${\rm s}^{-1}$ (Zavala et al. 2009), with many 
more low-velocity dispersion galaxies predicted by models. One potential explanation is that dark matter halos are strongly influenced by baryonic processes in this regime (Sawala et al. 2012).

Finally, in the bottom row of Fig. \ref{fig:rachel}, we show the cumulative
distributions of $v_{20}$ and $v_{\rm max}$.
In the bottom left panel, we show $v_{20}$ calculated from dark matter alone, in the middle panel $v_{20}$ including the effect of baryons and AC. These two figures can be used to read off predictions about the number density of galaxies with $v_{20}$ above a certain limit. In the bottom right panel of the same figure, we compare our cumulative $v_{\rm max}$ function to the analytic fit of Klypin et al. (2011) based on the Bolshoi simulation, which uses a WMAP7 cosmology (Komatsu et al. 2011). The agreement is reasonable given the different cosmology, with the MS-II simulation predicting more high mass haloes, as expected. Similar results have been obtained from a comparison between WMAP1 and WMAP7 velocity functions in Guo et al. (2013).

\section{Summary and Discussion}
Using the MS-II simulation, and particle data from the Milli-MSI, 
we have investigated the evolution of
the circular velocity of dark matter halos at 20 kpc, $v_{20}$.
Our main results are the following:
\begin{itemize}
\item For halos with mass $\log(M/M_{\odot}) > 12$ , the mass profile within 20--50 kpc changes only little since about $z \approx 4$, despite significant growth in halo
mass. Following the main progenitor history, we find almost no evolution in $v_{20}$ over this
time period.
\item As a consequence, the $v_{20}$ function of halos is almost in place at $z=4$.
\item By using $v_{20}$ as a proxy for the circular velocity 
of the galaxy, we can reproduce observations of the velocity dispersion of
massive galaxies by Bezanson et al. (2011) remarkably well. The agreement is
nearly perfect if we take into account the baryon content of halos inferred 
from abundance matching, and by considering the Blumenthal et al. (1986)
prescription for adiabatic contraction. The observations of Bezanson et al. (2011) 
are thus in excellent agreement with basic $\Lambda$CDM predictions on the evolution 
of structure in the Universe.

\end{itemize}
These results have interesting implications. First of all, we expect a significant
number of galaxies (more than one per $10^{6} h^{-3}{\rm Mpc}^{3}$) with circular velocities in excess of 450 km ${\rm s}^{-1}$ already 
at $z=4$. We also expect more than one per $10^{4.5}h^{-3}{\rm Mpc}^{3}$ with circular velocities in excess of 350 km ${\rm s}^{-1}$ up to $z=4$. Our results 
thus predict that high line widths should be found in high redshift galaxies with the Atacama Large Milimeter/Submilimeter Array (ALMA).

Second, we infer that the main progenitors of today's most massive galaxies
must have had high circular velocities already at high redshifts. This means that the circular velocity can be used to link massive galaxies with their progenitors and descendants.
It also helps to understand the fact that high redshift galaxies have higher  circular velocity 
at given dynamical mass than their local counterparts.
Recently, van de Sande et al. (2011) have measured a velocity dispersion of $\approx$ 300 km ${\rm s}^{-1}$ in 
a galaxy at $z=1.8$ with $M_{*}/M_{\odot} \approx  1.7 \cdot 10^{11}$, corresponding to a very high velocity dispersion
for its  dynamical mass compared to today's galaxies. As seen in Fig. \ref{fig:history}, the galaxy will continue to grow in halo mass
(and probably also in stellar mass) to $z=0$, but its circular velocity has likely almost reached its final value
by $z \approx 2$.  An even more extreme example is the extremely high velocity-disperson galaxy (510 km ${\rm s}^{-1}$) discovered
by van Dokkum et al. (2009). 

Finally, our result might offer an explanation for a puzzling
recent observational 
result. Wake et al. (2012a, b) have found that galaxy circular velocities
are an excellent predictor not only of the clustering properties of galaxies,
but also of galaxy
colours, specific star formation rates and ages, being superior to
galaxy stellar mass or surface mass density. As we have shown in
this work, the galaxy circular velocity, unlike the galaxy stellar mass, stays almost constant since early
times.  It thus might constitute a galaxy property that is more
fundamental than stellar mass,
retaining some memory of the initial conditions of the galaxy's formation.

\vspace*{-5mm}


\acknowledgements
 We thank the referee for helpful comments, which improved the paper. We acknowledge funding from ERC grant HIGHZ no. 227749.  SQL databases containing the simulation data from Millennium and Millennium-II 
    simulations are publicly released at 
    \texttt{http://www.mpa-garching.mpg.de/millennium}. The Millennium site was created as 
    part of the activities of the German
    Astrophysical Virtual Observatory. Tables for the velocity function of
halos as shown in this paper can be found at \texttt{http://www.strw.leidenuniv.nl/} \texttt{galaxyevolution/v20function/}.

\clearpage

\end{document}